\documentclass{ws-jnopm}
\begin{document}
\setcounter{page}{159}
\markboth{H.\ Susanto \& N.\ Karjanto}{Supratransmission Phenomena in Waveguide Arrays}

\catchline{}{}{}{}{}
\title{\vspace*{0.3cm}CALCULATED THRESHOLD OF SUPRATRANSMISSION PHENOMENA IN WAVEGUIDE ARRAYS WITH\\ SATURABLE NONLINEARITY
}

\author{H.\ SUSANTO}
\address{School of Mathematical Sciences, University of Nottingham\\
University Park, Nottingham, NG7 2RD, UK\\
hadi.susanto@nottingham.ac.uk}

\author{N.\ KARJANTO}

\address{School of Applied Mathematics\\ Faculty of Engineering and Computer Science\\ The University of Nottingham Malaysia Campus\\
Semenyih 43500, Selangor, Malaysia\\
natanael.karjanto@nottingham.edu.my}

\maketitle

\begin{history}
{\scriptsize \received{12 May 2008}}
\end{history}

\begin{abstract}
In this work, we consider a semi-infinite discrete nonlinear Schr\"odinger equation with saturable nonlinearity driven at one edge by a driving force. The equation models the dynamics of coupled photorefractive waveguide arrays. It has been reported that when the frequency of the driving force is in the forbidden band, energy can be transmitted along the lattices provided that the driving amplitude is above a critical value. This nonlinear tunneling is called supratransmission. Here, we explain the source of supratransmission using geometric illustrations. Approximations to the critical amplitude for supratransmission are presented as well.
\end{abstract}


\section{Introduction}

Mathematical models assuming nonlinear systems can lead to the emergence of new structures and nontrivial phenomena.\cite{scot03} An exotic phenomenon has been discovered recently by Geniet and Leon\cite{geni02} in a semi-infinite chain of coupled nonlinear oscillators driven at one edge by a time-periodic driving force. A plane wave scattered onto the medium by the periodic driving force should be totally reflected if its frequency falls in the system's band-gap. Yet, above a certain amplitude threshold, the incident wave was shown to generate a sequence of nonlinear modes propagating in the medium.\cite{geni02,capu01} This property is then called nonlinear supratransmission, which has been shown to be present in several different models,\cite{leon03} including nonlinear transmission in coupled optical waveguide arrays.\cite{Khomeriki}

In this work, we consider nonlinear supratransmissions in the discrete nonlinear Schr\"odinger (NLS) equation with saturable nonlinearity. The system, which is also called the discrete Vinetskii-Kukhtarev equation,\cite{Vinetskii} models the dynamics of the complex electric field envelope in waveguide arrays using photorefractive materials.\cite{step04} One particular motivation of choosing this system is its intriguing result that site-centered localized modes can exchange their linear stability with intersite-centered modes,\cite{hadz04} which leads to the existence of genuinely localized traveling waves.\cite{melv06} This waveguide array is then a good candidate for utilizing nonlinear supratransmissions to transmit binary information in discrete, semi-infinite system.\cite{maci1}$\!^-\!$\cite{maci3}  
It is significantly different from the regular discrete NLS equation with cubic nonlinearity, which has a short range of operating driving frequency for a quasi-traveling gap soliton.\cite{Khomeriki} The main purpose of this report is then to study this supratransmission, to understand its source and, if possible at all, to obtain the threshold amplitude numerically and analytically.

The report is presented as follows. In Section~\ref{numeric}, we consider numerically the discrete NLS equation with saturable nonlinearity as our model equation. We also present some simulations of the dynamics of the equation for different values of the driving amplitude, from which we illustrate the notion of supratransmission discussed in this paper. In Section~\ref{theoretic}, we analytically derive approximate expressions of the threshold amplitude for supratransmission. Finally, Section~\ref{conclude} gives a conclusion on supratransmission phenomena in a discrete NLS equation with saturable nonlinearity.

\section{Numerical Results} \label{numeric}

We consider the discrete equation\cite{step04}
\begin{equation}
\displaystyle i \frac{\partial \psi_n}{\partial z} + c\,(\psi_{n + 1} - 2\psi_{n} + \psi_{n - 1}) -
\gamma \frac{\psi_n}{1 + \left|\psi_n\right|^{2}} = 0,
\label{satunon}
\end{equation}
with $n = 2, 3, \dots, N,$ and driven at the boundary such that the governing equation at the site $n=1$ is given by
\begin{equation}
\displaystyle i \frac{\partial \psi_1}{\partial z} + c\,(\psi_2 - 2\psi_{1}+A e^{-i \Delta z}) -
\gamma \frac{\psi_1}{1 + \left|\psi_1\right|^{2}} = 0.
\label{boundary}
\end{equation}
The dependent variable $\psi_n$ then denotes the complex electric field envelope, the parameter $c$ is the coupling constant that characterizes
the tunneling coefficient, while the parameter $\gamma>0$ sets the strength of the nonlinearity. Without loss of generality, the coupling constant will be scaled to $c=1$. The variable $z$ represents the propagation distance variable, $A$ is the driving amplitude and $\Delta$ is the driving frequency. Following Ref.\ \refcite{Khomeriki}, the driving is turned on adiabatically by assuming the form $A\to A(1-\exp(-z/\tau))$ to avoid the appearance of an initial shock. In the following figures, we take $\tau=50$ and apply a linearly increasing damping to the final 10 sites to suppress edge reflection.

First, we need to recognize the allowed linear band of Eq.\ (\ref{satunon}). Substituting $\psi_n=\delta\exp\left({i(-\Delta z+kn)}\right)$, $|\delta|\ll1$, into the equation yields the dispersion relation $\Delta=\gamma-2c(\cos k-1)$ from which we obtain the linear band
\begin{equation*}
\gamma < \Delta < \gamma + 4c.
\end{equation*}

It is then expected that if the driving frequency is in the allowed band, any driving amplitude $A$ will excite all the sites. On the other hand, if the driving frequency is in the band-gap $\Delta<\gamma$ or $\Delta>\gamma+4c$, a small $A$ will only excite several neighboring sites. Yet, if the driving amplitude is large enough, then a train of moving discrete solitons can be released.\cite{Khomeriki} This flow of energy is the aforementioned supratransmission or the nonlinear forbidden band tunneling, and we call the minimum $A$ for supratransmission to occur as critical threshold $A_{\rm{thr}}$.
\begin{figure}[th]
\vspace*{-0.4cm}
\includegraphics[width = 6cm]{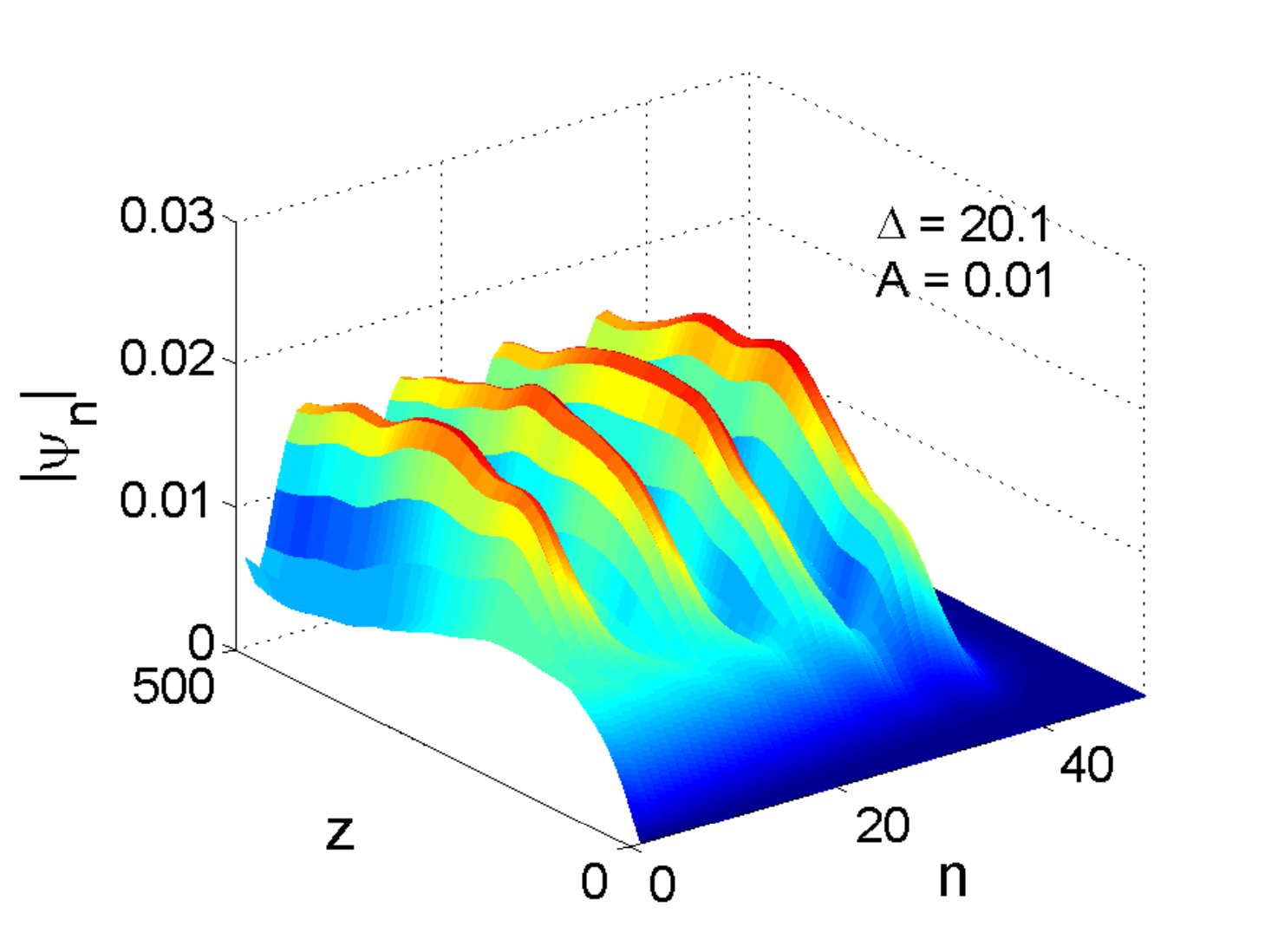}
\includegraphics[width = 6cm]{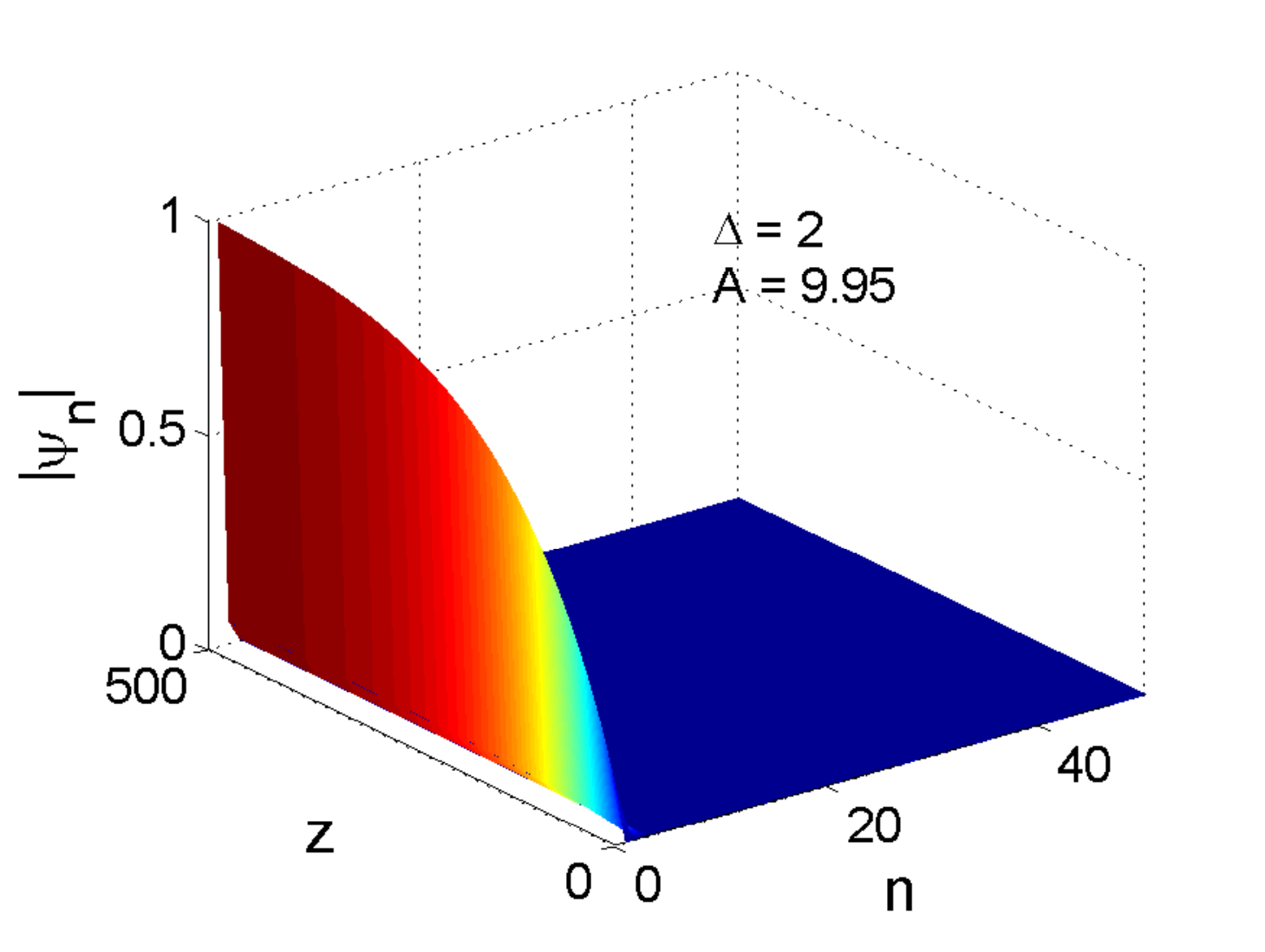}
\includegraphics[width = 6cm]{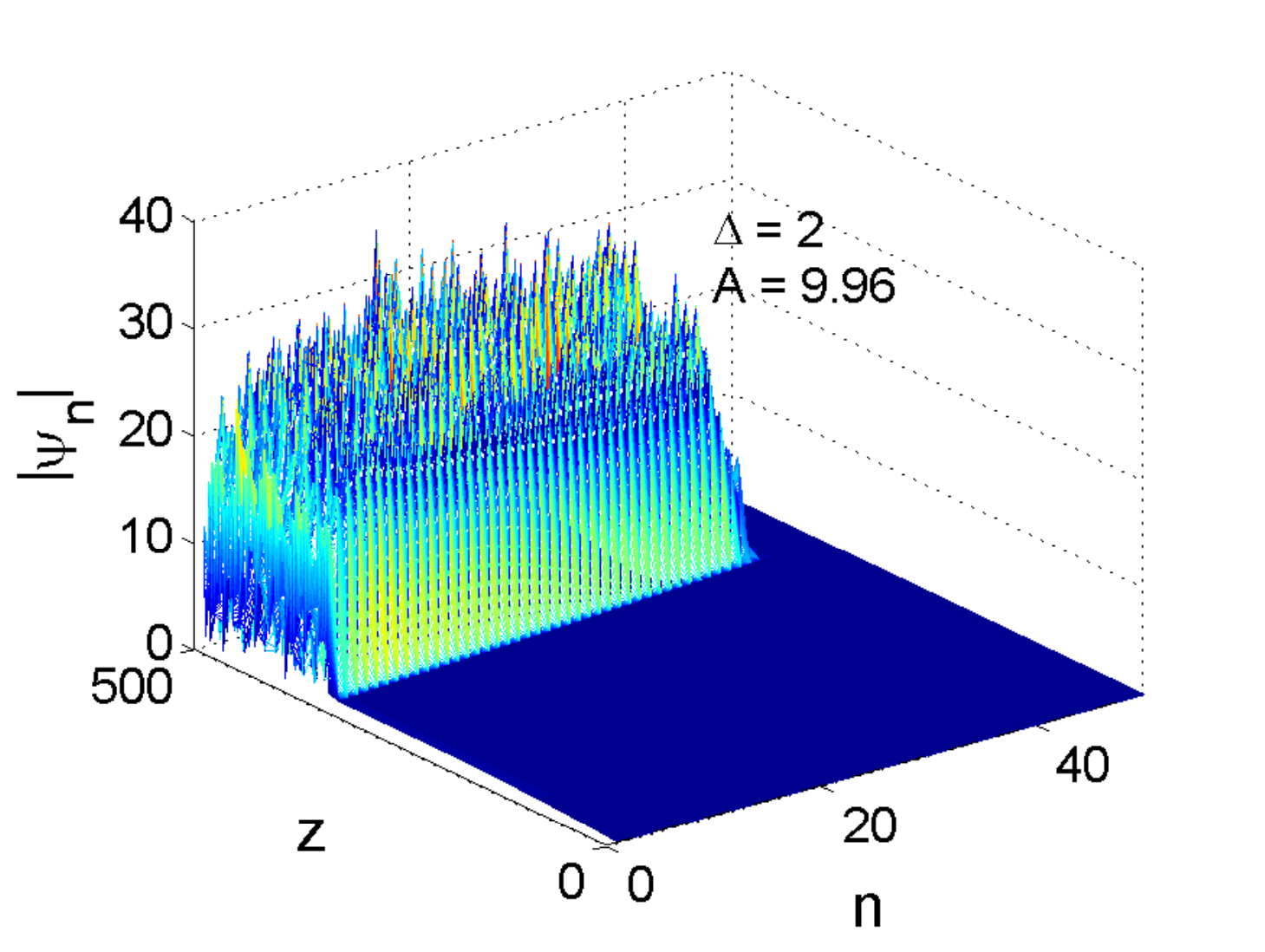}
\includegraphics[width = 6cm]{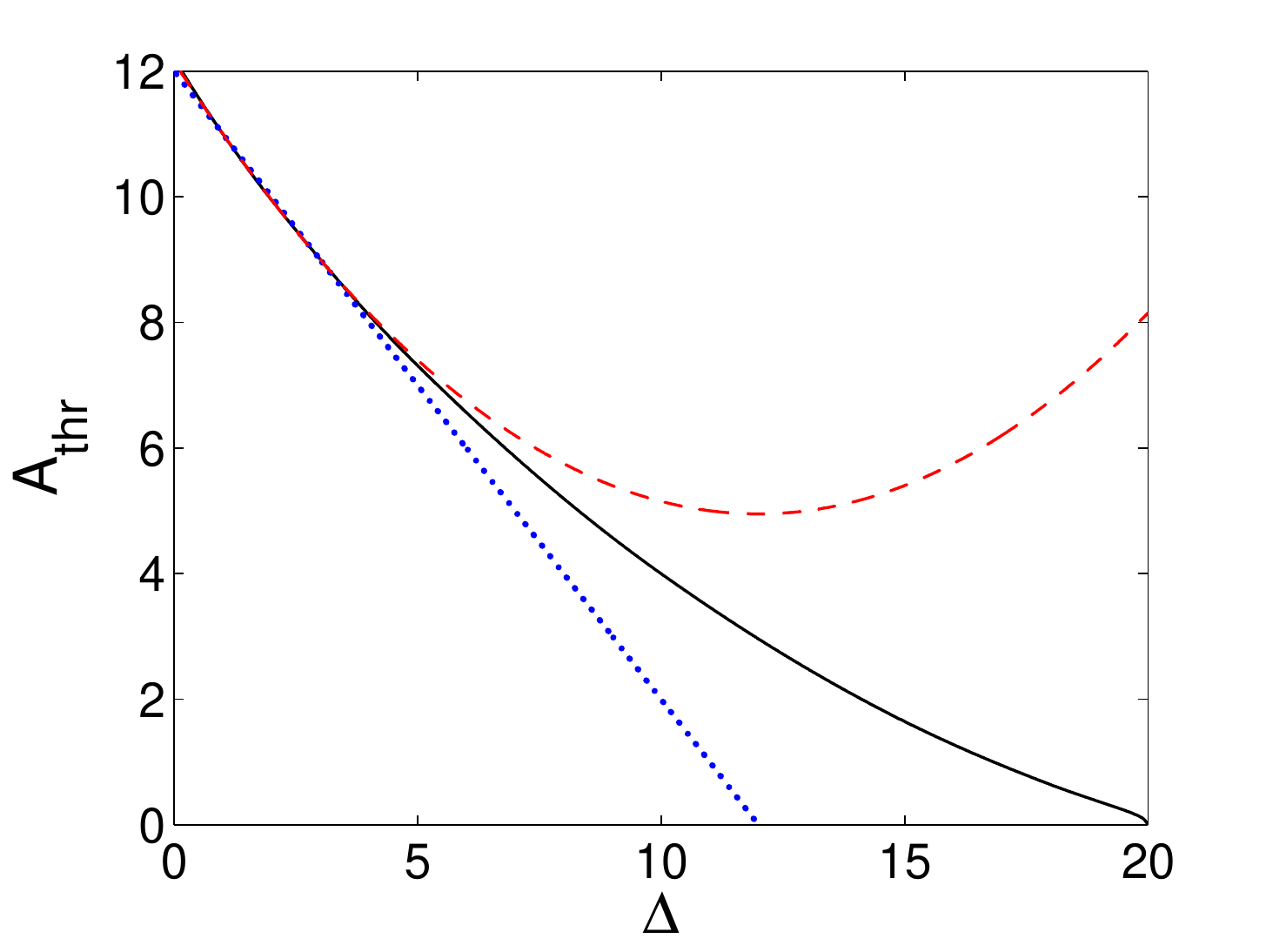}
\caption{When the driving frequency is in the allowed band, i.e.\ $\gamma<\Delta<\gamma+4c$, any driving amplitude $A$ will lead to
an energy transmission to remote sites. Shown in the top left panel is a three-dimensional plot of the dynamics of the NLS equation with saturable nonlinearity for the driving frequency $\Delta = 20.1$ and very small driving amplitude $A = 0.01$. The top right and the bottom left panels describe the dynamics of the equation for the driving frequency $\Delta = 2$ but the driving amplitudes are $A = 9.95$ (below the threshold value) and $A = 9.96$ (above the threshold value), respectively. The bottom right panel shows a comparison of the amplitude threshold $A_{\rm{thr}}$ (solid curve) and its approximations, i.e.\ Eq.\ (\ref{A0}) (dotted curve) and Eq.\ (\ref{A1}) (dashed curve), as a function of the driving frequency $\Delta$.}
\label{fig1}
\end{figure}

\vspace*{-0.4cm}
In this work, we consider the special case $\gamma\gg1$, which can be implemented experimentally.\cite{step04,step04_2} For illustrative purposes, throughout the paper we take $\gamma=20$. In Fig.\ \ref{fig1}, we present numerical simulations of the dynamics of Eqs.\ (\ref{satunon}) and (\ref{boundary}).

Shown in the top left panel is a three-dimensional plot of the time evolution in the allowed band corresponding to the NLS equation with saturable nonlinearity. Presented in the top right and bottom left panels are the dynamics of the equation when the driving frequency is in the band gap with the driving amplitude below and above the threshold value, respectively. Observe that the two plots have a completely different pattern even though the difference in the driving amplitude is very small. Summarized in the bottom right panel is the critical threshold amplitude $A_{\rm{thr}}$ as a function of the driving frequency $\Delta$. Shown in dotted and dashed curves are two approximations of $A_{\rm{thr}}$ that will be derived in the subsequent section. We observe that these approximations are in a good agreement with the numerical results when $\Delta$ is of $\mathcal{O}(1)$. 

\section{Theoretical Analysis} \label{theoretic}

Due to the application of the driving force at the boundary, it is then obvious that the complex-valued discrete function $\psi_n$ will have the same propagation constant, i.e.\ the frequency, as the driving frequency. Therefore, we seek stationary solutions to Eqs.\ (\ref{satunon}) and (\ref{boundary}) in the form of $\psi_n(z)=\phi_ne^{-i\Delta z}$, where $\phi_n$ is a real-valued function. Equations  (\ref{satunon}) and (\ref{boundary}) then give us
\begin{equation}
\begin{array}{lll}
\displaystyle \Delta\phi_1 + (\phi_{2} - 2\phi_{1} + A) - \gamma \frac{\phi_1}{1 + {\phi_1}^{2}} = 0,\\
\displaystyle \Delta\phi_n + (\phi_{n + 1} - 2\phi_{n} + \phi_{n - 1}) - \gamma \frac{\phi_n}{1 + {\phi_n}^{2}} = 0,\quad n=2,3,\dots
\end{array}
\label{static}
\end{equation}

\begin{figure}[th]
\centering \includegraphics[width = 6cm]{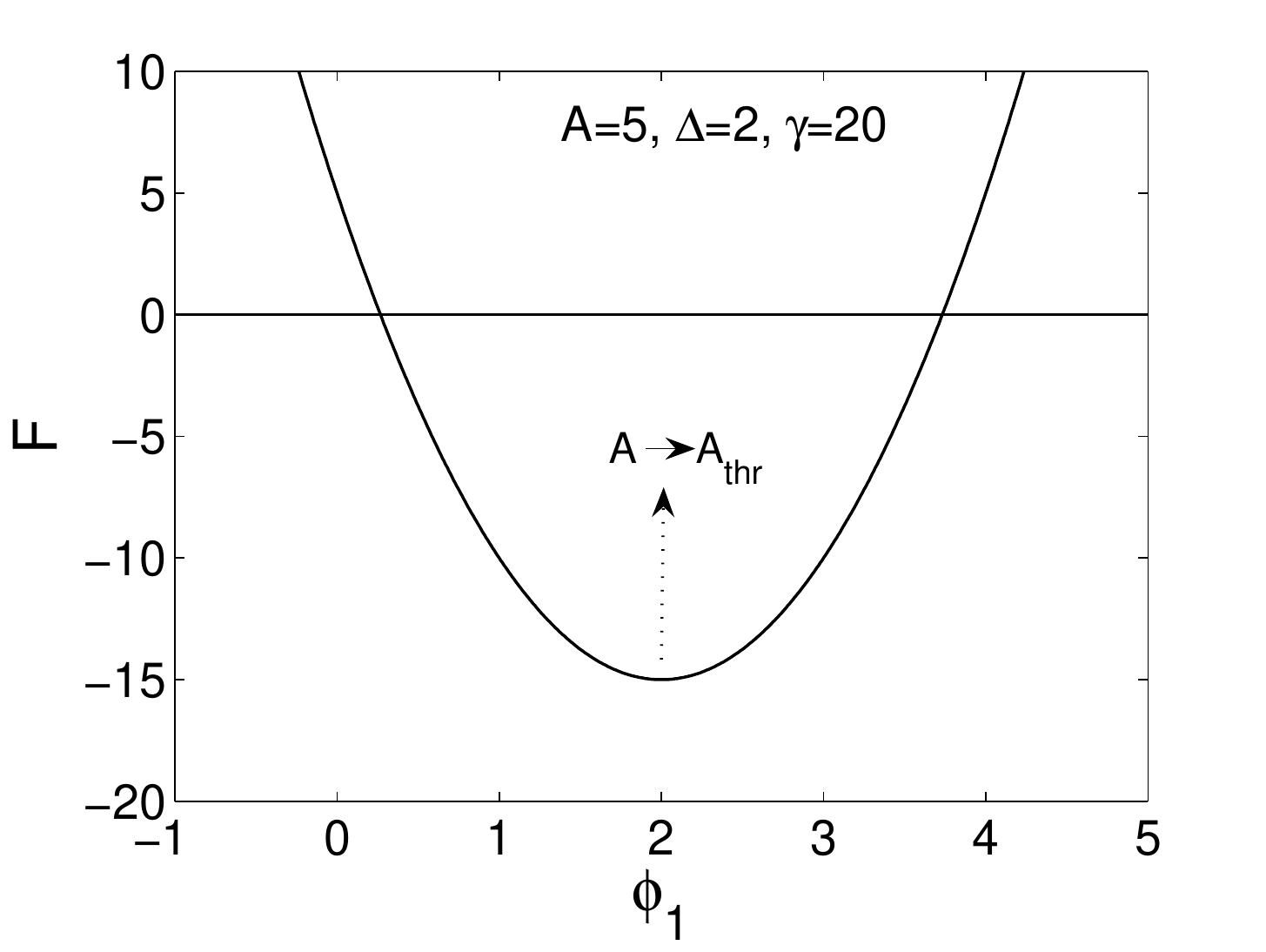}
\caption{The plot of $F$ [Eq. (\ref{static0})] as a function of $\phi_1$ for the driving amplitude $A = 5$ and the driving frequency $\Delta = 2$. The leading order of $F$ is, in general, a cubic polynomial, but taking these particular parameters reduces it to a quadratic polynomial with two real roots. For increasing $A$, the roots approach each other and eventually coincide when $A = \gamma/2 = 10$. For $A > \gamma/2$, there are no real roots, but instead imaginary ones.}
\label{fig2}
\end{figure}

As shown in Refs.\ \refcite{step04} and \refcite{step04_2}, when $\gamma\gg1$ and $\Delta\sim\mathcal{O}(1)$; the discrete system can be approximated by a few neighboring sites only. It is therefore justified in this limit case to approximate Eq.\ (\ref{static}) by
\begin{equation}
F:=\left(\Delta\phi_1 - 2\phi_{1} + A\right)\left(1 + {\phi_1}^{2}\right)  - \gamma {\phi_1}+\mathcal{O} \left(\frac{1}{\gamma^2} \right)= 0.
\label{static0}
\end{equation}
The roots of $F$ to the leading order would correspond to the so-called fast-decaying static solution to Eqs.\ (\ref{static}).\cite{Susanto} The plot of $F$ is presented in Fig.\ \ref{fig2}, where we zoom in on the region of the roots close to $\phi_1=0$.

As presented in Fig.\ \ref{fig2}, there are two particular roots of $F$ which are of interest to us. It is interesting to note that as $A$ increases, the solid curve of $F$ is shifted upward so that the two roots approach each other. Yet, there is a critical value of $A$ at which the roots coincide. If $A$ increases further, the roots disappear in a saddle-node bifurcation. Since these roots correspond to the fast-decaying static solution to (\ref{static}) (see the top right panel of Fig.\ \ref{fig1}), one can conclude that the critical $A$ at which the roots of $F$ disappear in a bifurcation also correspond to the threshold amplitude $A_{\rm{thr}}$ for supratransmission. Hence, we have naively shown that supratransmission exists because of a saddle-node bifurcation in the static solution to Eqs. (\ref{satunon}) and (\ref{boundary}). For a rigorous proof and detailed analysis, the reader is encouraged to consult Ref.\ \refcite{Susanto}.

A saddle-node bifurcation occurs when the local minimum of $F$ in Fig.~\ref{fig2} becomes a root of the function. Solving $F_{\phi_1}=0$ for $\phi_1$ yields $\phi_1=1+\mathcal{O}(1/\gamma)$. Substituting this value of $\phi_1$ to $F$ and solving it for $A$ then gives us the first two terms of the threshold amplitude,
\begin{equation}
A_{\rm{thr}} =\gamma/2 + (2 - \Delta) + \mathcal{O}\left(\frac{1}{\gamma}\right),
\label{A0}
\end{equation}
which approximates well the numerically obtained $A_{\rm{thr}}$ when $\Delta\sim\mathcal{O}(1)$, as is depicted by the dotted curve in the bottom right panel of Fig.\ \ref{fig1}. This approximation can be improved by including more higher order terms in the expansion of $F$, as illustrated in the following.

To calculate one order correction higher, the proper approximation of Eq.\ (\ref{static}), instead of (\ref{static0}), is now given by
\begin{eqnarray}
F_1:=\left(\Delta\phi_1 + \phi_2 - 2\phi_{1} + A\right)\left(1 + {\phi_1}^{2} \right) - \gamma {\phi_1} + \mathcal{O} \left(\frac{1}{\gamma^3} \right) &=& 0,	\label{static1a}\\
F_2:=\left(\Delta\phi_2 + \phi_1 - 2\phi_{2} \right)   \left(1 + {\phi_2}^{2} \right) - \gamma {\phi_2} + \mathcal{O} \left(\frac{1}{\gamma^3} \right) &=& 0.	\label{static1b}
\end{eqnarray}

By writing $\phi_1=1+\tilde{x}_1/\gamma+\mathcal{O}(1/\gamma^2)$, $\phi_2=\tilde{x}_2/\gamma+\mathcal{O}(1/\gamma^2)$, and $A_{\rm{thr}} \approx \gamma/2 + (2 - \Delta) + \tilde{A}/\gamma+ \mathcal{O}(1/\gamma^2)$, one will find from $F_2$ that $\tilde{x}_2=1$. Substituting this result into $F_1$ will yield
\[
\tilde{x}_1=4-2\Delta\pm2\sqrt{(\Delta-1)(\Delta-3)-\tilde{A}}.
\]

From this correction coefficient, it is indeed clear that the approximation (\ref{static1a})--(\ref{static1b}) [cf.\ (\ref{static0})] has two roots. The roots are real provided that $\tilde{A}\leq(\Delta-1)(\Delta-3)$. They collide and disappear in a saddle-node bifurcation when $\tilde{A}=(\Delta-1)(\Delta-3)$. Therefore, we next find that the first three terms of the threshold amplitude for supratransmission are given by:
\begin{equation}
A_{\rm{thr}} = \frac{\gamma}{2} + (2 - \Delta) + \frac{(\Delta-1)(\Delta-3)}{\gamma} + \mathcal{O} \left(\frac{1}{\gamma^2} \right). 	\label{A1}
\end{equation}

The plot of this approximation compared to the numerical results is also presented in Fig.\ \ref{fig1} (the dashed curve of the bottom right panel).

\section{Conclusion} \label{conclude}

To conclude, we have studied supratransmission phenomena in a discrete NLS equation with saturable nonlinearity and explained their source. We noticed that when the driving frequency is in the allowed band, any driving amplitude will lead to an energy transmission to remote sites. On the other hand, if the driving frequency is in the band-gap, only a small driving amplitude will excite several neighboring sites. We have shown the occurrence of supratransmission by comparing the dynamics of the equation by taking the driving amplitude below and above the critical threshold amplitude. We have also derived two approximate expressions to the threshold amplitude for supratransmission, which are good to the order of $\Delta\sim\mathcal{O}(1)$.

\section*{Acknowledgments}
The authors would like to thank Ardhasena Sopaheluwakan and Panayotis Kevrekidis for fruitful interactions.

\end{document}